# Possible Spectra of Cosmic Rays Accelerated in Extragalactic Sources

A.V. Uryson

Lebedev Physics Institute of Russsian Academy of Sciences

Moscow 117924

e-mail: Uryson@sci.lebedev.ru

**Abstract.** Spectra at the Earth of ultra-high energy cosmic rays (UHECR) from Seyfert nuclei with red shifts $z \leq 0.0092$ and from Blue Lacertae objects (BL Lac's) are calculated. The spectra are compared with the experimental data. It is shown that in nearby Seyferts the initial spectrum of the UHECRs is most likely the exponential one. UHECRs from BL Lac's may have the monoenergetic initial spectrum with the energy of $10^{21}$ eV to fit to the data. Spectra at the Earth derived in these two models are very similar if there is no Bl Lac's with red shifts $z<0.02$. The maximal energy in cosmic rays seemingly does not exceed $10^{21}$ eV. In calculations we used red shift distributions of the objects according to the recent (2001) catalogue of active nuclei by Veron-Cetty and Veron (see ref. in the text).

## 1. Introduction.

It is current opinion that cosmic rays with energies $E > 4 \cdot 10^{19}$ eV are of extragalactic origin. However their sources are not found out yet. Astrophysical objects of various kinds, as well as topological defects and unstable supermassive relic particles of cold dark matter along with gamma-bursts are considered as potential UHECR sources (see e.g. [1] and references therein). It is possible to identify UHECR sources in the first case if arrival directions of cosmic particles are known and if particles propagate in straight lines in the extragalactic space. I performed direct identification of UHECR sources in the papers [2, 3] and found that moderate Seyfert nuclei with red shifts $z \leq 0.0092$ along with BL Lac's were possible sources of UHECRs. BL Lac's were identified as possible sources also in [4]. The mechanism of particle acceleration in BL Lac's was supposed in [5] and the one in moderate active nuclei was suggested in my paper [6]. According to these results in BL Lac's particles attain the energy up to $10^{27}Z$ GeV, where $Z$ is the particle charge; taking into account possible losses in the sources the maximal energy is of $10^{21}Z$ GeV. In moderate Seyferts particles acquire the energy up to $8 \cdot 10^{20}$ eV. In extragalactic space particles interact with cosmic microwave background radiation and therefore inevitably loose energy [7, 8]. Energy losses of cosmic rays depend on its energies. Particles at different energies fly varying distances without significant losses. These distances were determined in [9, 10]. Red shifts $z \leq 0.0092$ of moderate Seyferts correspond to distances less than 40 Mpc (if Hubble constant is $H$=75 km/s Mpc) and these values of $z$ are in agreement with the results [9, 10]. BL Lac's identified as UHECR sources have red shifts up to $z$>1 [11]. Therefore it is unclear if the cosmic rays from BL Lac's can reach the Earth at the energy of $3 \cdot 10^{20}$ eV, that is the maximal energy of the observed cosmic ray particles [1].

In the present paper I calculate spectra at the Earth of particles coming from BL Lac's and from nearby Seyferts using red shift distributions of the sources in accordance with the data [11]. I suppose the initial spectra in the sources to be monoenergetic or exponential based on possible mechanisms of particle acceleration [5, 6]. The spectra obtained are compared with the measured one.

## 2. The model.

**2.1. Composition and initial spectra of UHECRs.** The composition of UHECRs at energies $4 \cdot 10^{19} - 3 \cdot 10^{20}$ eV is not clear yet. I suppose that UHECRs are particles rather than gamma-quanta [12].

According to [5] particles are accelerated in BL Lac's by an electric induction field in the vicinity of a supermassive black hole with the mass of $\sim 10^9 M_\odot$, $M_\odot$ is solar mass. The maximal particle energy is of $10^{27}Z$ GeV. If particles loose the energy in curvature emission in the source then the maximal energy is of $10^{21}Z$ GeV. Based on this mechanism of acceleration the spectrum of particles in BL Lac's may be monoenergetic with the initial energy $E_i=10^{27}$, $10^{21}$ eV. In moderate active nuclei particles may be accelerated by collisionless shocks [6]. In this case the initial spectrum of particles is exponential, $\sim E^{-\chi}$ [13]. Hence I adopt in calculations that the initial spectra are of these two kinds, the latter having $\chi$=2.6 or 3.0.

**2.2. Red shifts of cosmic ray sources.** We consider Seyfert nuclei with $z \leq 0.0092$ and BL Lac's from the catalogue [11] (with it these objects were identified as possible sources of UHECRs). Nuclei with declinations $\delta \geq -15^0$ and with defined red shifts were treated. According to [11] the nearest BL Lac's are at $z=0.02$. The statistics of nearby Seyferts is 105, the number of BL Lac's is 150. The red shift distributions of Seyferts and BL Lac's are shown in Fig.1,2. Each distribution is normalized to the total number of objects.

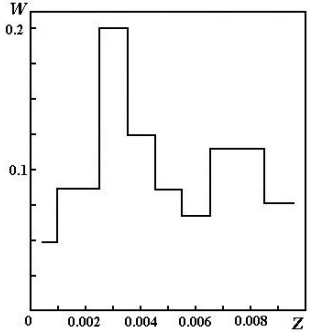 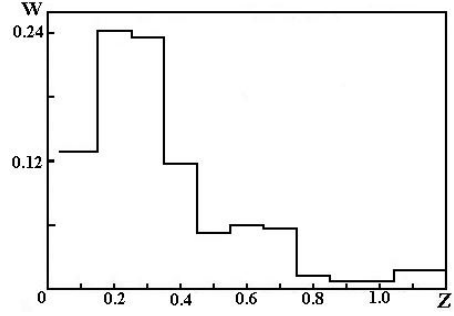

Fig.1. $z$-distribution of nearby Seyferts.   Fig.2. $z$-distribution of BL Lac's.

**2.3. Cosmological evolution.** Cosmic particles propagate distances from sources with large $z$, so I consider the cosmological evolution of the universe. Here Einstain — de Sitter cosmological model ($\Omega=1$) is adopted, in which time and red shift are related as

$$t=2/3H^{-1}(1+z)^{-3/2} \qquad (1)$$

and the distance to an object with a red shift $z$ is equal to

$$r=2/3cH^{-1}(1-(1+z)^{-3/2}) \text{ (Mpc)}, \qquad (2)$$

with $H=100$ km/(s Mpc). In the epoch with red shift $z$ the density of relic photons was $(1+z)^3$ larger and their energy was $(1+z)$ larger than at $z=0$ [13].

**2.4. Energy losses of cosmic particles.** In extragalactic space UHECR nuclei photodisintegrate in the infrared radiation field. Protons with energies $E>4\cdot10^{19}$ eV loose the energy in photoreactions with microwave background photons $p+\gamma\to N+\pi$ [13]. Energy losses due to pair production are inessential at $E>3\cdot10^{19}$ eV [14]. I consider here propagation of protons and do not treat nuclear fragmentation. This seems to be true in case of cosmic rays from BL Lac's because most of these sources is located at distances $r>200$ Mpc [11] and fragmentation lengths of ultra-high energy nuclei are ~10's Mpc [10,13,15]. For simplicity I suppose propagation of protons also for nearby Seyferts.

Proton propagation in the relic and in the infrared radiation fields was considered. The thermal-photon density spectrum is given by Planck distribution

$$n(\varepsilon)d\varepsilon=\varepsilon^2 d\varepsilon/(\pi^2(h/2\pi)^3 c^3(\exp(\varepsilon/kT)-1)), \qquad (3)$$

where the temperature is $T=2.7^0$ K, an average photon energy is $\langle\varepsilon\rangle\approx 6\cdot 10^{-4}$ eV, and a photon density is of $\langle n\rangle\approx 400$ см$^{-3}$. The photons of the high-energy "tail" of Planck distribution have an average energy $\langle\varepsilon_t\rangle\approx 1\cdot 10^{-3}$ eV and an average density of $\langle n_t\rangle\approx 42$ см$^{-3}$. The infrared radiation spectrum extends between $2\cdot 10^{-3}$ eV and 0.8 eV and is given by the numerical expression [10, 15]

$$n(\varepsilon)=7\cdot 10^{-5}\varepsilon^{-2.5} \text{ (см}^{-3}\text{ eV}^{-1}), \quad (4)$$

with an average energy $\langle\varepsilon_{IR}\rangle\approx 5.4\cdot 10^{-3}$ eV and an average density of $\langle n_{IR}\rangle\approx 2.28$ см$^{-3}$.

In addition to photopion losses protons loose energy due to red shift. The rate of the adiabatic losses is given by [13]

$$-dE/dt=H(1+z)^{3/2}E. \quad (5)$$

**2.5. Calculations.** In the proton rest system the photon energy is

$$\varepsilon^{*}=\gamma\varepsilon(1-\beta\cos\theta), \quad (6)$$

where $\gamma$ is proton Lorentz-factor, $\beta=(1-1/\gamma^2)^{1/2}$, $\theta$ is the angle between the momentum vectors of the photon and of the proton in the laboratory system. The cross section $\sigma$ and the inelasticity $K$ of the interaction depends on the photon energy $\varepsilon^*$. In calculations I use dependences $K(\varepsilon^*)$ and $\sigma(\varepsilon^*)$ from the paper [9]. The threshold energy for the production of pions is $\varepsilon^*_{th}\approx 145$ MeV and the threshold inelasticity is $K_{th}\approx 0.126$. In addition I adopt that $K\approx 0.5$ at $\varepsilon^*>2$ GeV, $\sigma\approx 0.07$ mbarn at $\varepsilon^*>5$ GeV.

On the first step of calculations the value of a red shift $z$ was simulated by Monte-Carlo technique in accordance with distributions shown in Fig.1,2. The distance to the source was calculated by (2). Then the proton energy $E$ and the angle $\theta$ in the laboratory system were simulated by Monte-Carlo method and the photon energy $\varepsilon^*$ was derived by (6). If it occured that $\varepsilon^*<\varepsilon^*_{th}$ the proton interacted with photons of the high-energy "tail" of Planck distribution. If the photon energy still was lower than the threshold energy the proton interacted with the infrared radiation. Values of the cross section $\sigma$ and of the inelasticity $K$ of the interaction were derived with the given value of $\varepsilon^*$. Then the proton mean free path was calculated

$$\langle\lambda\rangle=1/(\langle n\rangle\sigma),$$

and the value of the proton path $L$ was got by Monte-Carlo method. Next the value of the red shift $z_i$ at the epoch after the propagation of $L$ was calculated. Finally the energy of the proton after the propagation of $L$ was decreased by $E\cdot K$ due to the inelasticity and by $E(z-z_i)/(1+z)$ due to the adiabatic loss. The procedure was finished if the proton reached the Earth or if its energy decreased to less than $10^{19}$ eV.

### 3. Results

**3.1. The maximal energy of accelerated particles.** In this way I calculated average energies of protons that arrived from BL Lac's with initial energies $E_i=10^{27}$, $10^{21}$ eV. The statistics of simulated protons were $10^4$ in each case. Proton averaged energies at the Earth appeared to be $<E_f>\approx10^{24}$, $6\cdot10^{19}$ eV accordingly. The former value contradicts the experimental data, hence protons apparently do not attain the energy of $10^{27}$ eV in Bl Lac's. The initial energy $E_i=10^{21}$ eV seems to be suitable. This value is close to the maximal energy of $8\cdot10^{20}$ eV in moderate active nuclei [6]. For the further analysis we consider the UHECR spectra at the Earth.

**3.2. The spectrum of particles at the Earth.** The differential spectrum of UHECRs determined by different arrays is shown in Fig.3 taken from [1]. The spectrum has an irregular shape due to energy losses of particles in photopion processes. A flat component and a bump appear in the spectrum at $E>4\cdot10^{19}$ eV. The reason is that energetic particles are transferred in the region of lower energies while the lower is the energy of a particle the less are energetic losses [16, 17]. Spectra of particles from a single source and from sources distributed uniformly in the Universe were calculated in [18-20]. The energy at which the bump appears depends on distances from sources: if sources are near the bump appeares at higher energies. I analysed the shape of the measured spectrum in the range $10^{18} - 10^{20}$ eV in the paper [21].

The calculated spectra are also given in Fig.3. They are normalized to the measured spectrum (see the Fig.) The statistics of simulated protons are $10^5$ for each curve. It is difficult to make sure conclusions when comparing the calculated spectra with the measured one because of uncertainties in the experimental data. However one can see that two models seem to be ruled out: the model with the initial exponential spectrum in BL Lac's and the model with the initial monoenergetic spectrum in the nearby Seyfert nuclei (curves 1, 2a, 6). The initial exponential spectrum in the Seyferts is appropriate, but it is difficult to determine the value of the initial exponent χ, if χ=2.6 or 3.0 (curves 4, 5 in Fig.3) . The model of BL Lac's with the monoenergetic initial spectrum at $E_i=10^{21}$ eV also fits the measured spectrum (curve 3a).

Two models produce similar spectra at $E>10^{20}$ eV: the model with the initial monoenergetic spectrum in BL Lac's and the one with the exponential spectrum ($\chi=2.6$) in nearby Seyferts (curves 3a and 5). However if there is a 2 percent fraction of BL Lac's at red shifts $z=(0.01$-$0.02)$, in contrast to the minimal $z=0.02$ [2], these spectra become distinguishable (curves 3b and 5). In addition the exponential spectrum in Bl Lac's with $\chi=2.6$ maybe appropriate with this fraction (curve 2b).

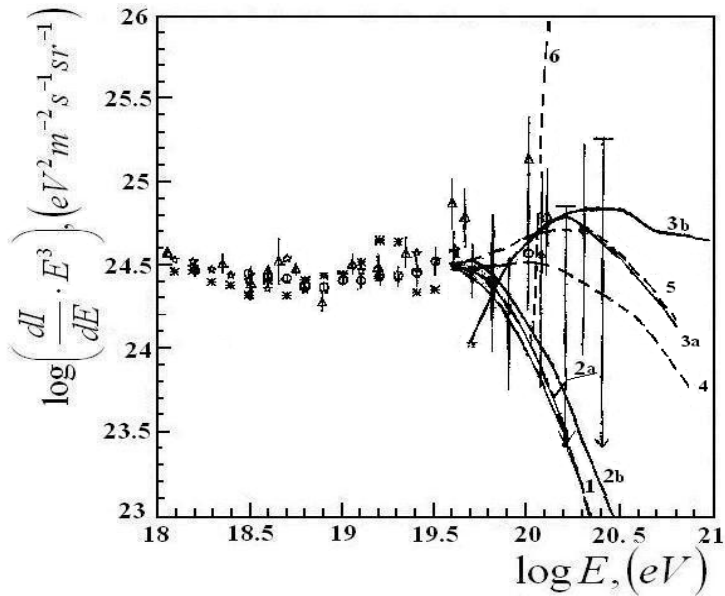

Fig.3. Differential energy spectrum of cosmic rays determined by the arrays Haverah Park, Yakutsk, Fly's Eye, AGASA, and Akeno 1-km$^2$ [2]. Curves are calculated UHECR spectra from nearby Seyferts (dashed lines) and from the BL Lac's (solid lines, label *a* denotes the model with the minimal redshift $z_{min}=0.02$, *b* is for the model with $z_{min}=0.01$): (1), (4) – the initial spectrum is exponential with $\chi=3.0$, (2a,b) and (5) - the initial spectrum is exponential with $\chi=2.6$, (3a,b) and (6) - the initial spectrum is monoenergetic.

**4. Discussion and conclusions.**

The maximal energy of cosmic rays seems to be of $10^{21}$ eV, no matter where they are accelerated, in Seyferts or in BL Lac's. This value is not far from the limits got in the models [22-24]: $\sim 10^{21}$ eV for particles accelerated in accretion disc, $\cong 10^{22}$ eV for UHECRs originated in gamma-ray bursts, and in the range $10^{21}$-$3\cdot 10^{21}$ eV if cosmic rays are produced in decays of metastable superheavy particles.

The spectrum from nearby ($z \leq 0.0092$) Seyfert nuclei may be exponential to fit the data. At present it is difficult to determine the value of the initial exponent of the spectrum as uncertainties in the experimental data are large. The calculated spectrum in case of Bl Lac's with the initial monoenergetic spectrum is also in agreement with the data. These spectra are similar if there is no Bl Lac's with red shifts $z<0.02$.

At energies below $10^{19}$ eV the spectrum could be formed by particles arising in a bulk of distant sources [13, 20]. At present the number of Seyfert galaxies is several thousands and the statistics of BL Lac's are several hundreds [2].

If the results obtained here are true the spectrum of UHECRs can be an additional test of the model [5] according to which superstrong electric fields are induced in the vicinity of a supermassive black hole.

The energy spectrum of cosmic rays at ultra-high energies will be determined with higher resolution and with larger statistics by ground arrays AGASA, HiRes, Pierre Auger, Telescope Array, EAS-1000, and by installations on board satellites [1, 25].

**Acknowledgment**

I would like to thank N.S.Kardashev for discussion of the model [5].

## ADDENDUM

**Abstract**. Some details of calculation that were omitted in astro-ph/0312618 are presented. Spectra of ultra high energy cosmic rays are shown with another value of Hubble constant $H$.

Some details of calculation were omitted in astro-ph/0312618: the distance $r$ is connected with the red shift $z$ by the relation

$$r = c\, z/H \text{ Mpc}, \qquad (1)$$

and is model independent to $z<0.4$. For $z>0.4$ the relation is

$$r = 2/3\, c/H\, (1-(1+z)^{-3/2}) \text{ Mpc}, \qquad (2)$$

in accordance with Einstein – de Sitter model with $\Omega=1$.

Here I present the spectra calculated with $H=75$ km/(s Mpc) using relation (2) for all values of $z$.

The calculated spectra and the experimental data from [2] are given in Fig.1. The statistics of simulated protons are $10^5$ for each curve. The spectra of UHECRs with the initial monoenergetic spectrum (curves 3, 6) are normalized at $E=5\cdot 10^{19}$ and $8\cdot 10^{19}$ eV respectively, other curves are normalized at $E=4\cdot 10^{19}$ eV. These values of $E$ were chosen to fit the data best. It is difficult to make sure conclusions when comparing the calculated spectra with the measured one because of uncertainties in the experimental data. However one can see that two models seem to be ruled out: the model with the initial exponential spectrum in BL Lac's if χ=3.0 and the model with the initial monoenergetic spectrum in the nearby Seyfert nuclei (curves 1, 6). The initial exponential spectrum in the Seyferts is appropriate, but it is difficult to determine the value of the initial exponent χ, if χ=2.6 or 3.0 (curves 4, 5 in Fig.3) . The model of BL Lac's with the monoenergetic initial spectrum at $E=10^{21}$ eV also fits the measured spectrum (curve 3).

Two models produce similar spectra at $E > 10^{20}$ eV: the model with the initial monoenergetic spectrum in BL Lac's and the one with the exponential spectrum χ=2.6 in nearby Seyferts (curves 3 and 5).

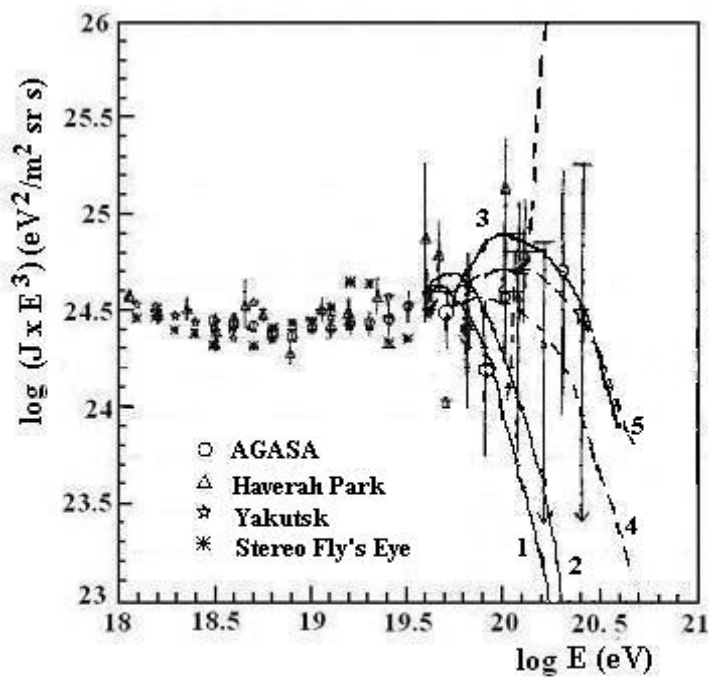

Fig.1. Differential energy spectrum of cosmic rays determined by the arrays Haverah Park, Yakutsk, Fly's Eye, AGASA, and Akeno 1-km² [2]. Curves are calculated UHECR spectra. Solid lines show calculated UHECR spectra from the BL Lac's: (1) – the initial spectrum is exponential with χ=3.0 , (2) – the initial spectrum is exponential with χ=2.6 , (3) – the initial spectrum is monoenergetic;  dashed lines show calculated UHECR spectra from nearby Seyferts: (4) - the initial spectrum is exponential with χ=3.0 , (5)- the initial spectrum is exponential with χ=2.6, and (6) - the initial spectrum is monoenergetic.